\documentstyle[prl,aps,preprint,tighten,floats,aps]{revtex}

\def\be{\begin{equation}}
\def\ee{\end{equation}}
\def\bear{\begin{eqnarray}}
\def\eear{\end{eqnarray}}
\def\beqn{\begin{eqnarray}}
\def\eeqn{\end{eqnarray}}

\def\beq{\begin{equation} }
\def\eeq{\end{equation} }

\def\ben{\begin{eqnarray} }
\def\een{\end{eqnarray} }

\def\mod#1{{\rm (mod~2)} }

\begin{document}
\draft
\preprint{\vbox{\baselineskip=12pt
\rightline{UPR-0831-T}
\vskip0.2truecm
\rightline{UM-TH/99-01}
\vskip0.2truecm
\rightline{hep-th/9903051}}}
\title{Blowing-Up the Four-Dimensional $Z_3$ Orientifold} 
\author{ Mirjam Cveti\v c$^\dagger$, Lisa Everett$^*$, Paul Langacker$^\dagger$, 
and Jing Wang$^\dagger$}
\address{$^\dagger$ Department of Physics and Astronomy \\ 
University of Pennsylvania, Philadelphia PA 19104-6396, USA\\
$^*$Department of Physics\\
University of Michigan, Ann Arbor, MI 48109  , USA}
\maketitle
\begin{abstract}
We study the  blowing-up of the   four-dimensional  $Z_3$ orientifold of
Angelantonj, Bianchi, Pradisi, Sagnotti and Stanev (ABPSS) by giving
nonzero vacuum expectation values (VEV's)  to the twisted sector
moduli blowing-up modes. The blowing-up procedure
induces a Fayet-Iliopoulos (FI) term for the ``anomalous'' $U(1)$, 
whose magnitude depends linearly on  the VEV's of the blowing-up modes.
To preserve the $N=1$ supersymmetry,  non-Abelian  matter
fields  are forced to acquire  nonzero VEV's,  thus breaking
(some of) the non-Abelian gauge structure and decoupling some of  the
matter fields. We  determine the form of the FI term, construct explicit
examples  of (non-Abelian) $D$ and $F$ flat directions,
 and determine the surviving gauge groups of the restabilized vacua. We
also determine the  mass spectra, for which the restabilization reduces
the number of families. 
 \end{abstract}
\newpage

\section{Introduction}
With the advent of the duality symmetries of M-theory~\cite{HullTownsend} a
new domain of string vacua (previously considered as strongly coupled)
have become accessible for study. Such dual four-dimensional string vacua
with $N=1$ supersymmetry   should provide a  fruitful domain for studying
novel phenomenological implications  of string theory. In particular, the
Type I string orientifolds~\cite{GimonPolchinski,orientifolds}  provide a
promising set of new string vacua, where  the techniques of the 
open-string theory allow for a quantitative study of the  gauge structure,
mass spectrum and (certain) couplings in the effective theory. 
The four-dimensional orientifolds~\cite{ABPSS,N1orientifolds}  with $N=1$
supersymmetry are thus particularly well suited for  phenomenological
studies.

One of the interesting features of the $N=1$ orientifolds is  that in
general  they contain a set of ``anomalous'' $U(1)$'s and the breaking of
these $U(1)$'s is inherently related to the blowing-up procedure. The
massless chiral superfields, formed from the
Neveu-Schwarz-Neveu-Schwarz (NS-NS) and 
Ramond-Ramond (R-R)  fields appearing in the twisted sector of type IIB
orientifolds,  are blowing-up moduli  whose nonzero vacuum expectation
values (VEV's)  correspond to the geometric smoothing-out (blowing-up)
of the orientifold singularities.
These fields play an instrumental role in the 
cancellation~\cite{Berkooz,U1anomaly}  of the
triangular gauge anomalies via the Green-Schwarz mechanism.
In addition, their coupling to the gauge superfields~\cite{douglasmoore}
contributes to the $D$ term of the
``anomalous'' $U(1)$, and its structure is fixed by the the anomaly
cancellation constraints. Thus, the Fayet-Iliopoulos (FI) term is
induced when the blowing-up modes acquire nonzero VEV's. To maintain the
anomalous  $U(1)$ $D$ flatness of the blown-up
orientifold,  additional matter fields in the theory have to acquire
nonzero VEV's subject to the constraint that the $F$ flatness and the
$D$ flatness of the rest of the gauge sector is maintained.  The generic
effect   of the blowing-up procedure and subsequent vacuum restabilization
is then the  spontaneous breaking of the
gauge symmetry as well as the decoupling of a number of matter fields in
the effective theory. In particular, the number of families in the
effective theory may be reduced.

It is instructive to contrast the blowing-up  and the 
accompanying vacuum restabilization of Type I orientifolds  with the 
blowing-up of perturbative heterotic string orbifolds. There, the
blowing-up of the orbifold singularities~\cite{cvetic}  and the
restabilization of vacuum due to the anomalous
$U(1)$~\cite{DSW,ADS,DK,AS} are somewhat disconnected and  the two
procedures are different from those on the Type
I side. For the $(2,2)$ orbifolds, i.e., those
with the spin and gauge connection identified  (analogs of such
constructions on the Type I side are non-existent), there is no
``anomalous''  $U(1)$ and the blowing-up
modes (twisted sector moduli)  are {\it charged} under the enhanced  gauge
symmetry that commutes with the discrete gauge
connection~\cite{stringsonorbifold,orbifolds}. (This is in contrast to those
of the Type 
I orientifolds which are total singlets.) Their nonzero VEV's fully break the
enhanced gauge symmetry that at the orbifold limit commutes with the discrete
spin connection and decouple some matter states; the procedure
geometrically corresponds to blowing-up the orbifold singularities
producing a smooth $(2,2)$ Calabi-Yau three-folds~\cite{cvetic}. On the
other hand  (asymmetric) orbifold  and free fermionic constructions with
only $(0,2)$ worldsheet symmetry in general possess an  anomalous $U(1)$.
In contrast with the Type I orientifolds such perturbative heterotic
string vacua have only one  anomalous $U(1)$, whose anomaly cancellation
is ensured by an universal Green-Schwarz mechanism, due to  an effective
Chern-Simons (CS) term, at the genus-one level, of the untwisted sector
two-form field to the gauge field strength\cite{DSW}. The dual of this
antisymmetric field-axion, along with the (untwisted sector) dilaton
field, form a scalar component 
of the chiral superfield, which  couples  universally to the gauge
sector of the theory. By supersymmetry the CS term is accompanied by a 
FI term (also at the genus-one level), which is proportional to the 
nonzero VEV of the dilaton field. Since this VEV  
determines the strength of the gauge coupling and thus it is necessarily
nonzero, these vacua necessarily have  nonzero  FI
terms~\cite{DSW,ADS,DK,AS}. The structure of this FI term is universal; it
is  completely fixed by the VEV of the dilaton and the trace of the
anomalous $U(1)$ charges. Therefore the existence of  ``anomalous''
$U(1)$ necessarily triggers the vacuum
restabilization. For recent work on the  systematic
classification of flat directions for a set of perturbative heterotic
string vacua see, e.g., Refs. \cite{cceel2,cceel3,cceelw}.

The crucial difference in the case of  the  Type I orientifolds  is  that 
 the blowing-up procedure and the restabilization of
vacuum are  now inherently connected. Since in general there are more than one
anomalous $U(1)$  the anomaly associated with each is cancelled by 
non-universal Green-Schwarz terms, which arise due to the CS coupling 
of the gauge field strengths to the twisted sector R-R two-form fields (which
is dual to the twisted sector R-R scalars -``axions''). Thus
each of the anomalous gauge group factors has a cancellation ensured by a
 specific  combination of twisted sector R-R 
axions. The non-universality of the Green-Schwarz mechanism via twisted R-R
sector axions is generic in  $N=1$ orientifold constructions\footnote{It is 
confirmed that at the orientifold limit there is no genus-one correction to
the FI term, calculated for  the ABPSS orientifold model in
Ref.~\cite{Poppitz}.}.  
Due to supersymmetry, these CS  terms are accompanied by
the corresponding FI terms, which involve a specific combination of the twisted
sector NS-NS fields-``blowing-up modes''.
 Consequently,  when these twisted sector dilatons acquire nonzero VEV's, 
the orbifold singularity, associated with  particular fixed points  where the
$D$ branes are located, is blown-up. This procedure  in turn induces 
 non-universal FI terms, which for each anomalous $U(1)$  is
proportional to a specific combination of the VEV's of the blowing-up
modes. The appearance of FI term then triggers the vacuum restabilization. 
 
The purpose of the present paper is to explore in a concrete way  the
effects of the blowing-up procedure for four-dimensional $N=1$
orientifolds.  The explicit  vacuum restabilization triggered by the
blowing-up  has to be carried out,  and  the surviving 
 gauge structure and  light  mass spectrum determined before phenomenological
 implications of the blown-up orientifolds can be addressed.
The goal is to carry out this procedure  explicitly for 
specific  blown-up orientifolds and to study its consequences.
(Unlike the four-dimensional $N=1$ orientifolds, 
the blowing-up procedure of  six-dimensional $N=1$ orientifolds 
is  better understood and related geometrically to  blowing-up of the ADE
singularities of  $K3$ surfaces \cite{douglasmoore}.)

While by now a large class of $N=1$ orientifolds have been
constructed~\cite{orientifolds,ABPSS}, only specific models (usually with
additional Wilson lines included) contain  matter fields which are
non-Abelian singlets. These  singlets  could be candidates for the
restabilization of the blown-up orientifold vacua, since in such  cases it
should be possible to achieve a systematic classification of flat
directions, very much along the lines developed for vacuum
restabilization of the perturbative heterotic string vacua  via non-Abelian
singlets~\cite{cceel2,cceel3}. 

On the other hand, a large class of (simpler) Type I orientifold constructions 
have no non-Abelian gauge
singlets, and thus the vacuum restabilization of  the blown-up orientifold
should necessarily proceed by giving VEV's to  non-Abelian matter fields.
In this case the  classification of  $D$ flat directions 
is  complicated by  non-Abelian $D$ flatness constraints, and 
a systematic approach to the study of vacuum restabilization 
is lacking.  Nevertheless, the  powerful connection 
between the holomorphic gauge-invariant monomials and the  $D$ flat
directions~\cite{monomialpapers} (which 
generalizes in the non-Abelian case to polynomials) facilitates the 
construction of non-Abelian $D$ flat
directions. Applications of this approach to the  vacuum restabilization of the
blown-up Type I orientifolds (without gauge singlets) will be  the focal point
of this paper. 

One of the immediate consequences of the vacuum restabilization with
non-Abelian fields is the breaking of the large non-Abelian gauge groups
down to smaller ones. From the phenomenological point of view, the
vacuum restabilization of the blown-up orientifold is one of a few ways to
achieve smaller gauge groups with reduced massless particle content. In that
sense, it could be viewed as complementary to other ``stringy'' methods in
orientifold construction, which involve spliting the 32 nine- (or five-)
branes among different fixed points of the orbifold, turning 
on the background NS-NS antisymmetric $B$-field, adding discrete
Wilson lines (see e.g., \cite{N1orientifolds}), etc.  

The paper is organized in the following way. In section II we describe the
spectra and the superpotential of the first 
four-dimensional orientifold with $N=1$ supersymmetry, constructed by
Angelantonj, 
Bianchi, Pradisi, Sagnotti and Stanev (ABPSS orientifold) \cite{ABPSS}. We
discuss the general procedure of anomaly cancellation and the generation of 
FI terms and explicitly write down the FI term for the anomalous
$U(1)$ of the ABPSS model. In section
III, we discuss the flatness conditions of the model in the restabilized
vacuum and the method of
classifying the $D$ flat directions. In section IV, we present classes of
$D$ flat directions that are also $F$ flat to all orders, and discuss the
consequences of the restabilized vacuum. In section V, we present the
conclusions.

\section{ABPSS model and anomaly cancellation}

\subsection{Massless spectrum  and  superpotential couplings of the ABPSS
orientifold}
 
We choose to analyze the ABPSS model. 
This  is  a $Z_3$ orientifold with the  gauge structure:  
$$SO(8)\times SU(12)\times U(1)$$
and a matter content of three copies of 
$$\psi^\alpha=(8,\overline{12})_{-1}, \ \ \ 
 \chi^\alpha=(1,{66})_{+2} \ \ \ 
\alpha=1,2,3,$$
which arise from  the open-string sector, due to the
strings stretching between the nine-branes located at the orientifold
singularities. 

In  the Type IIB orientifold (closed string) sector, in the NS-NS sector there
is the
gravity supermultiplet and the 36 (chiral) supermultiplets corresponding to
the $9$ untwisted (``toroidal'') and $27$ twisted (blowing-up)  sector moduli. 
The moduli  are  total  gauge singlets (unlike the twisted sector moduli of
the perturbative heterotic orbifolds), whose real and imaginary components
arise from the NS-NS and R-R sector, respectively. 
In particular,  the ABPSS orientifold has the
untwisted sector dilaton $S$, moduli singlets $T_{i}$ ($i=1,\cdots,
9$) and 27 twisted sector supermultiplets, out of which only one
participates in the blowing-up procedure, since all the nine-branes are sitting at the  fixed point of the $Z_{3}$ orbifold at the
origin. The twisted sectors ($k=1,2$) give the
NS-NS fields $\Phi_{k}$ and the R-R two-form fields $C_{k}^{(2)}$ (which
by duality are related to the twisted sector R-R axions $\Psi_k$). They
are constrained by the reality condition $\Phi_{1}=\Phi_{2}^{*}; ~
C_{1}^{(2)}=C_{2}^{(2)*}$.
In addition, the orientifold projection removes~\cite{DGM} the real
components of $\Phi_{k}$  and  $C_{k}^{(2)}$ ($\Psi_{k})$.

The renormalizable superpotential is of the form
\begin{equation}
W_3 = y \epsilon_{\alpha \beta \gamma} \psi^{\alpha a}_{i} \psi^{\beta b}_{i}
\chi^\gamma_{[a,b]}, 
\label{Weqn} 
\end{equation}
where $y$ is a constant; $\alpha, \beta, \gamma$ are family indices;
$i$ is an $SO(8)$ index; and $a,b$ are $SU(12)$ indices. 

In the following we shall address the nature of Chern-Simons (CS) terms in
orientifold models and derive the explicit expressions for the case of the
ABPSS orientifold. 

\subsection{Chern-Simons terms, Fayet-Iliopoulos terms, and anomaly
cancellation}

The $U(1)$ triangular gauge anomalies
are cancelled via the Green-Schwarz mechanism involving the exchange of twisted 
sector R-R  fields $\Psi_{k}$ (twisted axions) due to the CS
couplings~\cite{U1anomaly,douglasmoore}. For the four-dimensional Type I
orientifold the coupling takes the form:
\beq
\label{CS}
I_{CS}=\sum_{k} \int d^{4}x ~ C_{k} \wedge e^{F} = \sum_{k}\int d^{4}x
~C_{k}^{(2)} \wedge Tr(\gamma_{\theta^{k}})F + ...,
\eeq
where $C_{k}^{(2)}$ is the R-R 2-form in the $k^{\rm th}$ 
twisted sector; their  duals  are  the scalar  fields $\Psi_{k}$. $F$
schematically represents the gauge field strength  
of the  anomalous $U(1)$, associated with the $D$ brane located at
the orientifold singularity. The   matrix 
$\gamma_{\theta^{k}}$ describes the action of the orbifold group on the
Chan-Paton (CP) factors in the  $k^{\rm th}$ 
twisted sector. For the $Z_{N}$ orientifold, it takes the form
$\gamma_{\theta^{k}}=e^{-i 2\pi k V_{I} H_{I}}$ in the Cartan-Weyl basis, where
$V$ is a 16-dimensional real vector and $H_{I}$, ($I=1,..,16$) are the Cartan
generators of $SO(32)$ represented by 
tensor products of $ 2\times 2$ $\sigma_{3}$ submatrices. For the sake of
simplicity we confine ourselves to the case of $32$ branes  of the same type
located at a single fixed point.

The supersymmetric completion of the first term in eqn. (\ref{CS}) gives the
Fayet-Iliopoulos (FI) contribution to the action: 
\beq
\label{FI}
I_{FI}= \sum_{k}\int d^{4}x ~\Phi_{k} Tr(\gamma_{\theta^{k}}) D,
\eeq
where $D$ is the auxiliary component of the vector multiplet which contains
the gauge field $A_{\mu}$ of the anomalous $U(1)$. Hence, the FI term
is given by 
\beq
\label{FIp}
\xi_{FI}= \sum_{k} Tr(\gamma_{\theta^{k}} \lambda) \Phi_{k} , 
\eeq
where the sum is over twisted sectors, $\lambda$ is the Chan-Paton matrix
associated with the gauge boson of the anomalous $U(1)$. In the Cartan-Weyl
basis, it takes the form $\lambda_{i}=Q_{i} \cdot H$, where $Q_{i}$ is a
16-dimensional real vector. 
For the $Z_N$ orientifold, the spectrum of the $k^{\rm th}$-twisted
sector and the $(N-k)^{\rm th}$ twisted sector satisfy reality
conditions, e.g., $\Phi_k=\Phi_{(N-k)}^*$ and
$C^{(p)}_{k}=C^{(p)*}_{N-k}$. Furthermore, 
the orientifold projection projects out the real components of $\Phi_k$ and
$C_{k}^{(p)}$ (see e.g. \cite{DGM}). In
addition, the action of the orbifold group on the Chan-Paton indices 
 has the following  property:
\beq
Tr(\gamma_{\theta^k} \lambda) = [Tr(\gamma_{\theta^{(N-k)}} \lambda)]^{*}. 
\label{gamma}
\eeq
Then  the FI term takes the form:
\beq
\label{fieq2}
\xi_{FI}= 2\sum_{k=1}^{[\frac{N-1}{2}]} Re ( Tr(\gamma_{\theta^{k}} \lambda)
 \Phi_{k})=(-2)\sum_{k=1}^{[\frac{N-1}{2}]} Im[ Tr(\gamma_{\theta^{k}} \lambda) ]  Im(\Phi_{k}). 
\eeq
Note that due to the reality constraint, the sum is only over the first half
of the twisted sectors.

A similar argument applies to the coupling between the twisted sector R-R
scalar field $\Psi_{k}$ and the gauge field $A_\mu$, thus yielding the CS 
coupling of the type:
\beq
(-2) \int d^4x ~ \sum_{k=1}^{[\frac{N-1}{2}]}Im[ Tr(\gamma_{\theta^{k}} \lambda) ] \partial_{\mu} Im(\Psi_{k}) A^{\mu}.
\eeq
Thus the imaginary components of
$\Phi_{k}$ and $\Psi_{k}$ can be combined into the physical moduli $R_{k}$ of
the $Z_{N}$, $R_{k}= Im(\Phi_{k}) + i Im(\Psi_{k})$ in which $k$ goes from $1$ 
to $[\frac{N-1}{2}]$. 

For $Z_{3}$ orientifold models,  which have two twisted sectors,  the  
 reality condition on the twisted sector NS-NS scalars  reduces to
$\Phi_{1}=\Phi_{2}^{*}$ and the FI term (\ref{fieq2}) then takes the form
\beq
\label{FIeq3}
\xi_{FI}= -2 Im[ Tr(\gamma_{\theta^{1}} \lambda) ] Im(\Phi_{1}). 
\eeq
The physical moduli $R$ of the $Z_{3}$ of the single twisted sector is $R=
Im(\Phi_{1}) + i Im(\Psi_{1})$.    

A few comments are in order regarding the units appearing in front of the
CS term. The nine-brane CS  term  in 10 dimensions
has dimension one; it is proportional to
$\sqrt{\kappa_{10}}/g_{10}$~\cite{Polchinskibook}. 
Dimensionally reducing such a term to  four
dimensional effective theory, the prefactor gains  a volume factor 
$ V_{6}^{-1/4}\sqrt{\kappa_{4}}/g_{4}$. With the convention of assigning
VEV's of $Re(R)$ in terms of dimensionless quantities (just
as the convention for the dilaton field $S$), the prefactor of the FI
term (\ref{FIterm}) is  of dimension 2.  
However, since the CS-type couplings arise from the untwisted sector, 
they are absent in the four-dimensional theory of the $N=1$
orientifolds. Nevertheless we would like to argue that the dimensionful
parameters of the CS term associated with the twisted sector are of the
same structure.  
Such a  term should be calculated in perturbative open-string theory by
evaluating   the disk diagram for  two matter fields at the boundary and a
twisted field $C_k$
integrated over the bulk. While this calculation is technically involved due
to subtleties of the twisted sector fields\footnote{A related
calculation was given for six-dimensional untwisted sector fields in the
 Appendix of \cite{douglasmoore}.}, the dimensionful parameter
 of the resulting term is expected to have the same structure as that obtained
by a naive dimensional reduction.\footnote{In compactifications from $D=10$ to
$D=4$, the   case  in which the gauge groups arises from the
 five-brane world-volume theory, this relationship is modified by ratios of the
compactified 
five-brane world-volume    and the  volume  of the bulk. However in the
case of the  nine-branes (which fill up the full nine-dimensional spatial
part of the  ten-dimensional theory), the result depends only on the
volume of the six-dimensional space.} 

\noindent
{\it Gauge coupling corrections.} 
Analogously, one would like to determine the correction of the
twisted sector blowing-up modes to the gauge function:
\be
f=S+\delta f(R) \ .
\ee
Here $S$ is the (untwisted sector) dilaton  for the case of the nine-brane 
sector (it is the untwisted toroidal modulus $T$ in the case of the
five-brane sector.)
The  coupling of ${\rm Im}\delta f(R)F {\tilde F}$ could in principle appear
as the second order expansion of the  Chern-Simons term 
(\ref{CS}). In the case of $Z_{N}$ orientifold such a term takes 
the form: 
\beq
\sum_{k} \int d^4x ~ Tr(\gamma_{\theta_{k}} \lambda ^2) \Psi_{k} F {\tilde 
F}, 
\eeq
Thus when summed over the twisted sectors only the real component of $\Psi_k$ 
survives. However, it is projected out by the orientifold projection. Hence,
the term ${\rm Im}\delta f(R)F {\tilde F}$  
seems to be absent\footnote{One resolution to this problem may have to do
with   modifying the prescription that the sum is done only over the first
$[(N-1)/2]$
twisted sectors of the $Z_N$ orientifold. However, this would have to be
confirmed by explicit string calculations. We thank A. Uranga for
communication on that point.},  
indicating that $\delta f(R)=0 $, i.e., for $Z_{N}$ Type I orientifolds, there 
seems to be no gauge coupling correction due to the twisted sector moduli. 
Note that in contrast to Type I orientifold, Type II $Z_{N}$ orbifolds allow
such terms since the real
components of the twisted sector R-R fields are not projected out.

\noindent
{\it Anomaly cancellation.} 
The symmetry factors in the string amplitudes for a disc that involves
three gauge fields as external
legs can be determined by identifying the effective coupling of
the twisted-sector R-R fields to these gauge field strengths as arising from
the CS couplings of the type (\ref{CS}) (such
a string diagram can then be viewed as dominated by  the  
exchange of RR fields ~\cite{U1anomaly}). In
principle, the  prefactors of the amplitudes can be fixed by requiring the
exact cancellation of the gauge anomalies. In particular,
the amplitude of the scattering of a $U(1)_{i}$ gauge boson and two
non-Abelian $G_{j}$ gauge bosons with tree level exchange of a R-R scalar
is \cite{U1anomaly}\footnote{If the effective couplings
between the R-R fields and two non-Abelian gauge bosons are determined 
from the expansion of the CS term  of the type (\ref{CS}), and adding up
the contributions from all twisted sectors, the contributions from the $k^{\rm
th}$ and $(N-k)^{\rm th}$ twisted sector to such an effective term  add up to
zero, since the 
orientifold projection keeps only imaginary components of the $C_{k}^{(2)}$ forms.
However, the string amplitude that involves  the diagram with
three gauge bosons as external legs 
involves exchanges of R-R fields $\Psi_{k}$ in each twisted sector
separately. The total contributions to such a  string amplitude, summing
over the twisted sectors, would then 
yield a nonzero effective coupling that involves  two non-Abelian gauge bosons
 and the 
the anomalous $U(1)$ gauge boson, as quoted in \cite{U1anomaly}.}
  \beq
A_{ij}=\frac{i}{|P|} \sum_{k} C_{k} Tr(\gamma_{\theta^{k}} \lambda_{i}) Tr((\gamma_{\theta^{k}})^{-1}(\lambda_{j})^{2}),
\eeq
where $|P|$ is the order of the orientifold group and $k$ runs over the twisted
sectors. 
For a particular of the $Z_{N}$ orbifold, $C_{k}$ are given by~\cite{U1anomaly}
\begin{equation}
C_{k}=\Pi_{a=1}^{3}2 \sin{\pi k v_{a}}, 
\end{equation}   
where $v_{a}$ is the compact space twist vector. $A_{ij}$ cancels the usual
field theory triangular anomalies of the model~\cite{U1anomaly}. 
\\

\subsection{Fayet-Iliopoulos term and anomaly cancellation of the ABPSS model} 

In the ABPSS model, the CP matrix in the Cartan-Weyl
basis for the anomalous $U(1)$ gauge field
is $\lambda=diag\{I_{12}, -I_{12}, 0\times I_{8}\}$, in which $I_{n}$ is
the $n 
\times n$
identity matrix. The $\gamma$ matrices are given by 
\beq
\gamma_{\theta^{k}}=diag\{e^{ik \theta}I_{12}, e^{-ik \theta}I_{12}, 
I_{8}\},
\eeq
where, $\theta=\frac{2 \pi}{3}$. 

The FI term of the anomalous $U(1)$, which arises from  the
supersymmetric completion of the Chern-Simons
couplings between the twisted sector R-R fields and the gauge
fields~\cite{douglasmoore}, is given by eqn.(\ref{FIp}) with $\lambda$ and
$\gamma_{\theta^{k}}$ of the ABPSS model:
\begin{equation}
\xi_{FI}= - 2\times 12 \times 2\sin(\frac{2 \pi}{3})~ Re(R) = -24 \sqrt{3}
~Re(R),
\label{FIterm}
\end{equation}
where $R$ is the twist sector moduli field. 
$\xi_{FI}$ modifies the usual $U(1)$ $D$ term as
\begin{equation}
D\to D+\xi_{FI} .
\end{equation}
Therefore, the FI term is proportional to the VEV of the real component of the
twisted moduli $R$.

The  coupling of ${\rm Im}\delta f(R)F {\tilde F}$ in the case of ABPSS
orientifold could take the form: 
\beq
\sum_{k=1,2} \int d^4x ~ Tr(\gamma_{\theta_{k}} \lambda ^2) \Psi_{k} F {\tilde 
F} = 2\times 12 \times \cos{\frac{2 \pi}{3}} \times 2 \times Re(\Psi_{1}). 
\eeq
Since the real component of $\Psi_1$ is projected out by the orientifold
projection, the term ${\rm Im}\delta f(R)F {\tilde F}$ seems to be absent (as
discussed on general grounds in the previous subsections). 

In the ABPSS model, the $U(1)^3$,  $U(1)\times SO(8)^{2}$, and $U(1)\times
SU(12)^{2}$ anomalies are
\beq
\label{U(1)anomalies}
(432, ~-36, ~18), 
\eeq
respectively. 

Since the compact space twist vector for $Z_{3}$ is
$v=\frac{1}{3}(1,1,-2)$, one obtains $C_{1}=-C_{2}=-3 \sqrt{3}$. 
Thus, the scattering amplitudes $A_{ij}$, with the $\gamma$ and $\lambda$
matrices presented previously, are \cite{U1anomaly}  
\begin{equation}
A_{U(1),~(U(1), SO(8), SU(12))} = \frac{1}{3}2(-3\sqrt{3})12\frac{\sqrt{3}}{2}
(24(\frac{1}{2}), -1, \frac{1}{2})=(-432, 36, -18),
\end{equation} 
which cancel the $U(1)$ anomalies in eqn.(\ref{U(1)anomalies}).  \\


\section{Anomalous $U(1)$ and Vacuum Restabilization}

The appearance of the FI term for the anomalous $U(1)$ due to the blowing-up
procedure 
requires the well known vacuum restablization procedure to preserve
supersymmetry at the string scale. Certain fields that are charged
under the anomalous $U(1)$ are triggered to acquire nonzero vacuum
expectation values (VEV's) that cancel the FI $D$
term, subject to the constraints that they are both $D$ flat with respect
to the other gauge groups and $F$ flat, leading to a consistent
``restabilized" string vacuum.
As a consequence, some fields become massive (depending on the size of the
FI term, which generally sets the scale of the VEV's,  they will either
decouple or remain in the low energy theory).
In addition, the rank of the gauge
group is usually reduced as well as the number of families.

In previous work \cite{cceel2}, techniques were developed to
construct the moduli space of the flat directions for models
with an anomalous $U(1)$ systematically. The method utilizes the
one to one correspondence of $D$ flat directions (under both the
non-anomalous Abelian gauge groups and the non-Abelian gauge groups) with
holomorphic gauge-invariant polynomials built out of the chiral
fields in the model. For simplicity,
the flat direction analysis in \cite{cceel2} considered only the non-Abelian
singlet fields in the model, in which case the flat directions correspond to
monomials (HIM's). In particular, the superbasis, which is the set of
the one-dimensional (i.e., that depend on one free VEV) HIM's of the
model, can be constructed.
Every $D$ flat direction can be expressed as a
product of the elements in the superbasis, such that the positivity of the
VEV-squares of the fields can be satisfied automatically. 
For example, if the $l^{th}$ HIM for the flat direction is
$P_l = \Pi_p \Phi_p^{n^l_p}$, then the fields $\Phi_p$ have
VEV's
\beq | \langle \Phi_p \rangle |^2 = \sum_l n^l_p |v_l|^2,
\label{product} \eeq
where $v_l$ is the VEV corresponding to $P_l$. The phase of the
$| \langle \Phi_p \rangle |$ can be chosen for convenience.

The HIM's of the superbasis are then classified according to the sign of
their contribution to $D_{A}$, the $D$ term of the anomalous $U(1)$.
Since the FI term for the anomalous $U(1)$ has to be cancelled by the
VEV's of certain fields in the model, 
the sign of the FI term is crucial. 
To ensure the $D_{A}$ flatness constraint,  the required HIM's should
necessarily contain one or more elements that are opposite in sign to
that of $\xi_{FI}$.

The constraints of $F$ flatness require that $\langle \partial W /
\partial \Phi_{p} \rangle =0$ and $\langle W \rangle =0$ for all of
the massless superfields $\Phi_{p}$ in the model. 
Further consideration of these conditions demonstrates that there
are two types of dangerous terms which can lift a given $D$ flat
direction.
The first class of terms, which we denote as the $W_{A}$ terms,
are formed solely of the fields that are in the $D$ flat direction. 
Gauge invariance dictates that if such a term can be constructed, it can
appear in the superpotential raised to any positive power. In this case, 
for the $D$ flat direction to remain $F$ flat to all orders in the
superpotential, string selection rules must conspire to forbid the
infinite number of $W_A$ terms, which can be difficult to prove in
general.  We choose to adopt a conservative strategy and do not consider
$D$ flat directions for which $W_A$ terms can appear (with the recognition
that in doing so, we may be neglecting possible flat directions which are in
fact $F$ flat to all orders).

The other type of dangerous terms, which we denote as the $W_{B}$ terms, 
are linear in an additional massless superfield which is not in the flat
direction (i.e., which has zero VEV), such that $\langle W \rangle =0$ but
$\langle \partial W / 
\partial \Psi \rangle$ may be nonzero.  In this case, gauge invariance
constrains the number of $W_B$ terms to be finite, and an explicit string
calculation can be performed to determine if such terms are in fact
present in the superpotential.  Thus, the flat direction can be proven to
be $F$ flat to all orders if all such $W_B$ terms vanish.  
It is also possible that in certain
cases the contributions to
the $F$ term $\partial W / \partial \Psi$ from different $W_B$ terms linear in
the same field $\Psi$ (which is not in the flat direction) could be
arranged to cancel for appropriate magnitudes and signs of the VEV's
of the fields involved.

In the present model, the $D$ flat directions necessarily involve
non-Abelian fields due to the matter content. In principle, the problem
could be simplified significantly if only one 
component for each superfield which is charged under $SU(12)$ and/or
$SO(8)$ is nonzero, such that only diagonal generators would be involved
in $D$ terms for the non-Abelian gauge
groups. In this case, the problem is similar to that of the Abelian case
({\it Abelian-like}), and the techniques 
developed in \cite{cceel2} and described above can be directly applied. 

The one to one correspondence between holomorphic gauge invariant 
polynomials (HIP's) and non-anomalous $D$ flat directions \cite{monomialpapers}
provides a powerful way of searching for a more general class of $D$ flat
directions with  
non-Abelian fields. We first construct a gauge invariant polynomial from the
non-Abelian fields, which is a sum of monomials involving the components 
of the fields. Then one monomial term defines a $D$ flat direction. Each field
in the monomial will have the same magnitude of the VEV (or times $\sqrt{n_p}$ if
the component field is raised to the $n_p$ power). The $D$ flatness
constraints for both diagonal and off-diagonal generators of the
non-Abelian gauge group are automatically
satisfied. Other flat directions are gauge rotations of the direction
corresponding to a single monomial. They are equivalent to a product of
monomials from the same HIP. Each
monomial introduces the same magnitude of the VEV for each component present
(or $\sqrt{n_{p}}$), but the phases of the VEV's are dictated
by the gauge rotation.

One can also consider higher dimensional $D$ flat directions
(with more than one independent VEV), formed as products of
other HIP's. The flat directions correspond to products
of monomials from each of the HIP's, each with its own VEV.
Such products
often have a reduced surviving gauge symmetry and massless particle
content. They are sometimes $F$ flat (due to cancellations) for specific
ratios of the VEV's and choices of phases, even though the directions
corresponding to a single HIP are not.
For {\it overlapping} polynomials, which are products involving common
multiplets, there is a flat direction in which
the common multiplets
have the same nonzero component (or involve components not
connected by any single gauge generator) for each of the   
monomial factors, which avoids  nonzero $D$ terms for off-diagonal
generators\footnote{Gauge or family rotations of that flat
direction may have additional nonzero components of the same
multiplets. The $D$ flatness of the off-diagonal generators
then occurs by cancellations.}.
The VEV's of the nonzero components in a
product are given by
an expression analogous to (\ref{product}).
For products of
{\it non-overlapping} HIP's (i.e., with no multiplets in common), 
the corresponding monomials may have different orientations
in the internal symmetry space.

\section{Flat Directions}

The $F$ flatness conditions of the ABPSS model are
\beq
\epsilon_{\alpha \beta \gamma} \psi_{i}^{\alpha a} \psi_{i}^{\beta b}=0; 
\eeq

\beq
\epsilon_{\alpha \beta \gamma} \psi_{i}^{\beta b} \chi^{\gamma}_{[a,b]}=0.
\eeq
The $D$ flatness condition for $SO(8)$ is 
\beq
D^{I}=\sum_{\alpha, a}\sum_{i,j}(\psi_{i}^{\alpha a \dagger} T^{I, ij}
\psi_{j}^{\alpha a })=0,
\eeq
where $T^{I}$ are generators of the vector representation of $SO(8)$ and
$I=1,..,28$. For $SU(12)$, 
\beq D^J = \sum_{\alpha,i} \sum_{a,b} \psi_i^{\alpha a \dagger}
\hat{T}_{ab}^J \psi_i^{\alpha b}
+ \sum_{\alpha} \sum_{a,b,c} \left( \chi^\alpha_{[a,c]}\right)^\dagger
T^J_{ab} \chi^\alpha_{[b,c]},  \eeq
where $T^J$ ($\hat{T}^J \equiv - T^{JT}$) are the generator matrices for
the fundamental (anti-fundamental) representation of
$SU(12)$ and $J=1,..,143$. The $D$ flatness condition for the anomalous $U(1)$ is
given 
by
\beq
D_{A}=-\sum_{\alpha, i, a} \psi_{i}^{\alpha a \dagger}\psi_{i}^{\alpha a}+2
\sum_{\alpha, a<b}\chi^{\alpha \dagger}_{[a,b]} \chi^{\alpha}_{[a,b]}+\xi_{FI}.
\eeq 

Since there are no non-Abelian singlet fields in the model, the $D$ flat
directions are necessarily formed of 
superfields which transform nontrivially under the non-Abelian gauge
groups. 
Due to the form 
of the superpotential and the number of families, we have not been able to
 construct a $D$ flat direction involving only one component per 
multiplet, namely, the Abelian-like solution with only the diagonal generators
of $SU(12)$ involved. 

We thus concentrate on the holomorphic gauge invariant polynomial method as a
more general and powerful tool. We construct gauge invariant (under $SO(8)
\times SU(12)$) combinations of
fields $\psi^{\alpha a}_{i}$ and/or $\chi^\gamma_{[a,b]}$. Any monomial from
the HIP involving particular components of the fields $\psi$ and/or $\chi$
is a one-dimensional $D$ flat direction. Other $D$ flat directions can be
constructed from products of these one-dimensional directions.
We then check the $F$ flatness constraints.    

We have found classes of flat directions involving $\chi$ only,
exploiting the fact that the totally antisymmetric product
$\chi^6$ is an $SU(12)$ singlet. Different combinations of
the family indices for the six $\chi$ fields, or products
of such polynomials, correspond to different residual symmetries
and spectra after vacuum restabilization.
 The $\chi$ fields have 
$U(1)_{A}$ charge $+2$, so that these directions have $D_A = 0$
for  $\xi_{FI} < 0$, i.e., for positive $Re(R)$ in (\ref{FIterm}).
The VEV's  of the components of the fields are proportional to
$\xi_{FI}$, so they interpolate smoothly to the limit $\xi_{FI} = 0$.

The one-dimensional flat directions take the form 
(up to a gauge rotation)
\beq
\chi^6=\chi^{\alpha_1}_{[1,2]}\chi^{\alpha_2}_{[3,4]} \chi^{\alpha_3}_{[5,
6]}\chi^{\alpha_4}_{[7, 8]} \chi^{\alpha_5}_{[9, 10]} \chi^{\alpha_6}_{[11,12]},
\eeq
where the family indices $\alpha_i$ take the values 1, 2, or 3. This 
monomial is a singlet under $SO(8)$, with anomalous $U(1)$ charge $+12$.

In addition to $SO(8)$, which obviously remains unbroken by this class of flat 
directions, the remaining unbroken gauge symmetries from $SU(12)$ clearly
contain $SU(2)^{6}$,
where each $SU(2)$ corresponds to the indices of one of the $\chi$ fields in
the 
flat direction. However, since there are only three possible values of the
six $\alpha_i$, these are not all the unbroken generators
of the original $SU(12)$. There are additional off-diagonal generators which
remain unbroken such that the remaining gauge symmetry is $Sp(2l)\times Sp(2m) 
\times Sp(2n)$\footnote{Note that in the orientifold limit, appearance of
$Sp(2k)$ groups is usually associated with existence of
five-branes. Interestingly, the ABPSS orientifold has only nine-branes, yet
the 
blowing-up procedure introduces $Sp(2k)$ groups.}, where $l$, $m$, and $n$ are the number of occurrences of the
direction 1, 2, and 3, respectively, and $l+m+n=6$. (Of course, $Sp(2) \sim
SU(2)$ and the $Sp(2k)$ factor is absent for $k=0$, where $k$ is $l$, $m$ or
$n$.) In Appendix A, an example with two $\chi$ fields is given to illustrate
the survival of the $Sp(4)$ gauge group.

There are also flat directions which are arbitrary superpositions of
directions with different family indices  but the same
$SU(12)$ indices. These are equivalent to each of the six factors 
having an independent direction in family space;
i.e., 
$\alpha_i$ is promoted to a vector in the three-dimensional
family space, with each component having an arbitrary phase. Thus,
$\chi_{[a_i,b_i]}^{\alpha_i} \rightarrow 
\chi_{[a_i,b_i]}^{1} \chi_{[a_i,b_i]}^{2} \chi_{[a_i,b_i]}^{3}$,
where $[a_i,b_i]=[1,2],[3,4],[5,6],[7,8],[9,10]$, or $[11,12]$,
and the VEV's of three components in the family space satisfy 
$|v_{[a_i,b_i]}^{1}|^{2}+|v_{[a_i,b_i]}^{2}|^{2}+|v_{[a_i,b_i]}^{3}|^{2}
=|v|^2$, where $v$ is the VEV determined from $\xi_{FI}$.
The generic residual symmetry is $SU(2)^6$, except for the
special directions for which $k$ of the $\alpha_i$ are aligned,
in which case $SU(2)^k \rightarrow Sp(2k)$.


The space of the flat directions of this class is a subspace of the $3\times
66$ dimensional complex space of $\chi^{\alpha}_{[a, b]}$. At a generic
point (the  vectors $\alpha_{i}$ in the family space are different from
each other), other 
flat directions can be constructed by family and phase transformations
classified by $U(3)/U(2)$ for 
each of the six factors $\chi_{[a_{i}, b_{i}]}^{\alpha_{i}}$. Of the five
generators of $U(3)/U(2)$, one is from  
$SU(12)\times U(1)_{A}$ and the other four correspond to moduli. Hence, the
generic points with $6$ arbitrary $\alpha_{i}$ form a $12$-dimensional
complex moduli space. The points in moduli
space that involve permutating the $\alpha_{i}$ and associated phases are
equivalent by discrete gauge transformations, such as the one which
maps $\chi^{\alpha_1}_{[1,2]} \chi^{\alpha_2}_{[3,4]}$ to
$\chi^{\alpha_1}_{[3,4]} \chi^{\alpha_2}_{[1,2]}$. 
The $\chi$ spectrum for a generic point includes 126 massive states associated
with the spontaneous breakdown of $SU(12)\times U(1)_{A}$ to $SU(2)^{6}$
(the imaginary parts are the absorbed Goldstone bosons and the real parts
are the massive scalar partners). In addition, there are $3\times 66 -
126=72$ massless complex $\chi$ states, $12$ of which are associated with
the moduli. At the special points in the moduli space where the vectors
in the family space $\alpha_{i}$ are aligned, the gauge symmetries are
enhanced. There are correspondingly more massless states and fewer
associated with symmetry breaking.   

To determine the spectrum for the $\psi$ fields, consider
any of the six factors of $\chi ^6$.
Without loss of generality we can choose axes in family space such that 
$\langle \chi^{\alpha}_{[a,b]} \rangle = v$ for a specific
$\alpha =$ 1, 2, or 3, with the other VEV's vanishing. From the superpotential 
\beq
W_{3} \rightarrow  2yv (\psi_{i}^{\beta a} \psi_{i}^{\gamma b} - 
\psi_{i}^{\gamma a} \psi_{i}^{\beta b}),
\eeq
(where $\beta \neq \gamma \neq \alpha$) one has that $\psi_{i}^{\beta a}$,
$\psi_{i}^{\gamma b}$, $\psi_{i}^{\gamma a}$ and $\psi_{i}^{\beta b}$ become
massive ($\sim 2 y|v|$), while $\psi_{i}^{\alpha a/b}$ remain massless.  
Similarly, two of the three families become massive for each of the 12
values of the $SU(12)$ index. ($SO(8)$ remains unbroken.) In the
special case of $l=m=0$, $n=6$, for example, $\psi_{i}^{3 a}$, ($a=1..12$)
remain massless in the restabilized vacuum. Hence, three apparent
families are reduced to one. This suggests that
it may be worthwhile to consider models with more than three apparent families.

The $\chi^6$ flat directions are $F$ flat to all orders, since  $SO(8) \times 
U(1)_{A}$ invariance requires at least two factors of $\psi$ for
each term in the superpotential.
The form of the effective superpotential (after restabilization) for the 
massless states  depends on the specific flat direction. $U(1)_A$ invariance
implies that $n_\psi = 2 n_\chi$ for an allowed term in the original
superpotential, where $n_\psi$ and $n_\chi$ are respectively the
number of factors of $\psi$ and $\chi$. Effective cubic
terms must therefore result from surviving terms in the original 
superpotential $W_3$ in (\ref{Weqn}). Such terms survive for 
points in the moduli space except for the maximal symmetry points
like $(l,m,n)=(0,0,6)$.  There could conceivably be four-dimensional
effective couplings from original non-renormalizable couplings of the type
$\psi^4 \langle 
\chi^2 \rangle$.  However, these would have to be of different form than
$W_{3}^2$.

The generic $\chi^6$ flat directions that we have discussed so far
are technically of the overlapping type; i.e., they involve products
of monomials in which more than one component of the same
field is allowed to have a VEV, as occurs, for example, if
$\alpha_1$ and $\alpha_2$, associated  with $\chi^{\alpha_1}_{[1,2]}$ and
$\chi^{\alpha_2}_{[3,4]}$ are not
orthogonal. For this class of flat direction this presents no
difficulty: no $SU(12)$ generator connects $\chi^{1}_{[1,2]}$ with
$\chi^{1}_{[3,4]}$, for example, and no off-diagonal $D$ terms are induced.
For this to work, it is necessary for the $SU(12)$ indices to be
the same in each monomial, i.e., each has the same $SU(12)$ orientation.

There is another class of $\chi^6$ flat directions involving
non-overlapping polynomials, such as $(\chi^1)^6 (\chi^2)^6$,
or $(\chi^1)^6 (\chi^2)^6 (\chi^3)^6$. Since they
are non-overlapping, there is no need for the $SU(12)$ indices to be
the same in each factor. These directions therefore allow
even more breaking of the $SU(12)$, although to maintain the
non-overlapping character there is much less freedom for rotations in the
family indices\footnote{There are also hybrid flat directions, involving
family rotations for some of the $\chi$ factors and $SU(12)$ rotations
for others.}. There are many possibilities for the relative $SU(12)$
orientations of the $(\chi^\alpha)^6$ factors, with different
implications for the physics of the associated restabilized vacua. We will
simply use two examples given in Appendix B to illustrate the 
complexity of the physics associated with this class of flat directions.


There are other classes of directions which are $D$ flat
with respect to $SO(8) \times SU(12)$, of the generic form 
$\psi^{12}$ or $(\psi^{12}) \cdots (\psi^{12})$.
Such directions have negative
$U(1)_{A}$ charges, and would yield $D_A = 0$ for blown-up
constructions with $\xi_{FI} > 0$. These would correspond to
$Re(R) < 0$ in (\ref{FIterm}), and would not have a canonical geometric
interpretation. For such directions,
each $(\psi^{12})$ factor is totally antisymmetric in the $SU(12)$
indices, while the $SO(8)$ indices are contracted in pairs.
The latter can be within  the same $(\psi^{12})$ factor, in which case
they must have different family indices, or can connect
different $(\psi^{12})$ factors. However, we have not found any examples
of this class  which are also $F$ flat\footnote{We have also not
found any $F$ flat examples for generalizations involving
overlapping polynomials in which the monomials have different combinations of
$SU(12)$, $SO(8)$, and family indices. These generically induce $D$ terms for
the off-diagonal generators, but these can often be cancelled by appropriate
choices of relative signs for the fields in a monomial.}.

Similarly, we have not found any non-trivial solutions involving both the 
$\psi$ and $\chi$ fields in the flat direction. Clearly $\psi^{\alpha a}_{i} 
\psi^{\beta b}_{i} \chi^\gamma_{[a,b]}$ is not $F$ flat, and we have not found
combinations involving different families, etc., which are both $F$ and $D$
flat\footnote{In \cite{ABPSS}, a form of directions involving both $\psi$ and
$\chi$ which satisfied $F$ flatness was given. However, no examples which were
$D$ flat were presented, nor was their existence proved. The examples
presented in  this paper are both $D$ and $F$ flat, and are outside of
their class.}.
(Of course, the orientifold point with $Re(R) = 0$  has $\xi_{FI} = 0$, and
the trivial solution with no nonzero VEV is $D$ and $F$ flat.)

\section{Conclusions}
The aim of this paper has been to address explicitly the   blowing-up
procedure of the four-dimensional Type I  orientifolds with $N=1$
supersymmetry. 
The specific analysis was done for the  Type I $Z_3$ orientifold constructed
by Angelantonj, Bianchi, Pradisi, Sagnotti and Stanev (ABPSS orientifold)
\cite{ABPSS}. We chose it in part due to its simplicity (it involves only
nine-branes), and in part due to its potential  phenomenological
implications since the model contains three families.  The goals  were
two-fold: (I) we identified the fields participating in the blowing-up
procedure, and determined the structure of the induced Fayet-Iliopoulos
(FI) term, and (II) we provided a  detailed analysis of
the flat directions, surviving gauge symmetry, and light particle spectra
after the subsequent vacuum restabilization of the blown-up orientifold.

(I) For the $Z_3$ ABPSS orientifold, only one twisted sector
blowing-up mode  $R$  participates in the blowing-up procedure and
triggers the subsequent  vacuum  restabilization.  
It is associated with one out of the 27 fixed points, chosen at the
origin where the nine-branes are located.   We showed that the  real part 
$Re(R)=\phi$ of this blowing-up mode (which arises in  the NS-NS  twisted
Type IIB sector) contributes to the FI term.  The nonzero 
vacuum expectation value (VEV) of this field which, based on the
geometrical interpretation of the blowing-up procedure should be taken
to have positive sign, fixes the magnitude and the sign of the FI
term.  On the other hand, the  Chern-Simons (CS) term, which plays
an instrumental role in the anomaly cancellation procedure and is related
to the FI term by supersymmetry,  is proportional to the  imaginary  part
$ Im(R)=\psi$ of the blowing-up mode (which arises from the R-R twisted
sector).  We also noted that  the second order 
expansion of this CS term~\cite{douglasmoore}, which contributes to the 
imaginary part of the gauge coupling corrections, does not seem to be
present (the part of the  R-R sector two-form  field
that would contribute to this coupling is projected out by the orientifold
projection.).

(II) Due to the absence of non-Abelian singlet fields in the spectrum, the
subsequent vacuum restabilization necessarily involves  {\it non-Abelian
fields}, thus complicating the analysis.
We employed the powerful connection between holomorphic gauge
invariant polynomials and flat directions~\cite{monomialpapers},
which enabled us to classify the $D$ and $F$  flat directions
of the blown-up orientifold in detail.  We found that the $D$ and $F$
flat directions of this model are associated with the general set of $\sim
\chi^6$ monomials, in which $\chi$ transforms  as  $ (1,66)_{+2}$ under
$SO(8)\times SU(12)\times U(1)$.  We were unable to
find $D$ and $F$ flat directions for the gauge invariant monomials of the
type $\sim \psi^{12}$, in which $\psi$ transforms as
$(8,{\overline{12}})_{-1}$ under $SO(8)\times SU(12)\times U(1)$.   We
were also unable to find hybrid flat directions with both the $\psi$ and
$\chi$ fields involved. 
Incidentally, only  the $\sim \chi^6$  $D$ flat directions of this model
have the correct sign of the $U(1)$ charges to cancel the FI term
with the positive value of the blowing-up mode.
 
The generic point in the moduli space  of $D$ and $F$ flat directions
associated with the  $\sim \chi^6$ monomials is
specified by 12 complex moduli (fields that can acquire free VEV's),
breaks the gauge group down to $SU(2)^6\times SO(8)$, and leaves only one 
family of the $\psi$ multiplets massless. At special points of moduli
space, for example where $k$ of the family indices $\alpha$ in the
monomials are aligned, the unbroken gauge group  $SU(2)^k$ is enhanced to
$Sp(2k)$.

We conclude with a number of remarks that in view of the analysis
presented in this paper may be more general and could apply to a
larger class of Type I blown-up orientifolds.
In particular, a preliminary investigation of other orientifold models  
with non-Abelian singlets, indicates that the vacuum
restabilization procedure still necessarily involves non-Abelian
fields~\cite{celpw}. A large set of $Z_N\times Z_M$ orientifolds (e.g.,
\cite{orientifolds}) have only the fundamental and antisymmetric tensor
representations of the gauge group in the light particle spectrum (i.e.,
those  of  the $\psi$- and $\chi$-type, respectively).
It is conceivable that ingeneral the $\chi$ type fields play an instrumental
role in the vacuum restabilization of this class of blown-up orientifolds.

Our analysis demonstrates that the particular blown-up orientifold
addressed in this paper is unlikely to have interesting  phenomenological
implications,  since neither the surviving gauge group, generically
$SU(2)^6\times SO(8)$,  nor the particle content, with generically only
one family remaining light, are phenomenologically viable.
Nevertheless, the approach sets the stage for further systematic analysis
of other blown-up orientifolds, which may uncover potentially
phenomenologically interesting Type I string vacua.

\acknowledgments
This work was supported in part by U.S. Department of Energy Grant
No. DOE-EY-76-02-3071.  We would like to than M. Douglas, G. Moore, E.
Poppitz, and Z. Kakushadze for useful correspondence and  A. Uranga for
correspondence and discussions. We would also like to thank M.
Pl\"{u}macher for helpful discussions.   
\newpage

\appendix
\section{Enhanced symmetries.} 

As an example, take two $\chi$ fields from the same family,
and assume the components $\chi^{3}_{[1,2]}$ and $\chi^{3}_{[3,4]}$ have
nonzero 
and equal VEV's. Label the $SU(12)$ generators by $T^{a}_{b}$, where
\beq
T^{a}_{b} \chi_{[c,d]} = \delta ^{a}_{c} \chi_{[b,d]} +\delta^{a}_{d}
\chi_{[c,b]}.
\eeq
Then, the unbroken generators in the $SU(4)$ subgroup are $(T_{1}^{2},
T_{2}^{1}, T_{1}^{1}-T_{2}^{2})$ and $(T_{3}^{4}, T_{4}^{3},
T_{3}^{3}-T_{4}^{4})$, which correspond to $SU(2)^2$, as well as
$T_{3}^{1}-T_{2}^{4}, T_{1}^{4}+T_{3}^{2},
T_{1}^{3}-T_{4}^{2}, T_{4}^{1}+T_{2}^{3}$. It is straightforward to show that
the remaining gauge group is $Sp(4)$. Generalizing to the case with
$k$ factors of $\chi$ fields aligned in their family indices, $3k$
generators 
remain unbroken from the obvious $SU(2)^k$ subgroup. In addition, there are 4
unbroken generators for each pair of $\chi$ fields. Therefore, the total
number of unbroken generators is $3k+4 k(k-1)/2=k(2k+1)$, and the unbroken
gauge group can be shown to be $Sp(2k)$. 
There are six broken diagonal generators, which are the five $SU(12)$
generators not in $SU(2)^6$ and that of the anomalous $U(1)$.  The 
off-diagonal generators which connect factors from different
families are also broken, as are the linear combinations of the
off-diagonal generators that are orthogonal to the generators in the
extensions of $SU(2)^k$ to $Sp(2k)$. \\

\section{Directions involving non-overlapping polynomials.}

We will simply illustrate with two examples involving an $SU(4)$ subgroup
of $SU(12)$ and two $\chi$ fields. 

As we have seen in Appendix A,
the single monomial $\chi^1_{[1,2]} \chi^1_{[3,4]}$ breaks  $SU(4)$
to $Sp(4)$, while $\chi^1_{[1,2]} \chi^2_{[3,4]}$ leaves $SU(2)^2$ unbroken.
As an example of a non-overlapping polynomial direction, consider
$(\chi^1_{[1,2]} \chi^1_{[3,4]}) $ $(\chi^2_{[1,4]} \chi^2_{[2,3]})$,
where the first (second) pair of fields have VEVs $v_1$ ($v_2$).
Although this direction breaks each individual $SU(4)$ generator,
there are six unbroken linear combinations. These
are $t_1 \equiv T^1_2 -T^3_4$, $t_2 \equiv
T^1_4 + T^3_2$, and their Hermitian
conjugates, as well as the Hermitian generators $h_1 \equiv
T^1_1 - T^2_2 + T^3_3 - T^4_4$ and $h_2 \equiv i(T^2_4 + T^1_3 - T^3_1 - 
T^4_2)$. $h_1$ and $h_2$ commute with each other and the $t_i$. The
six surviving generators correspond to an unbroken $SU(2)^2$, with the
canonical $SU(2)$ generators given by linear combinations of the $t_i$ and
the $h_i$. This class of directions involves $5$ real moduli. There are $8$ 
massless complex $\chi$ fields associated with the $SU(4)$ subgroup. Two
families of $\psi$ states become massive and one remains massless. 

An extension of this example utilizes all three families, i.e.,
$(\chi^1_{[1,2]}\chi^1_{[3,4]})$ $ (\chi^2_{[1,4]} \chi^2_{[2,3]})
$ $(\chi^3_{[1,3]} \chi^3_{[4,2]})$, with VEVs $v_i, i=1,2,3$ for the
three pairs. In this case, there are three
Hermitian generators, $h_2$, $i(t_1 - t_1^\dagger)$, and
$i (t_2 - t_2^\dagger)$, of a surviving $SU(2)$. There are $7$ real moduli for 
these examples, and $5$ massless complex $\chi$ fields. All three families of
$\psi$'s become massive.

\def\B#1#2#3{\/ {\bf B#1} (19#2) #3}
\def\NPB#1#2#3{{\it Nucl.\ Phys.}\/ {\bf B#1} (19#2) #3}
\def\PLB#1#2#3{{\it Phys.\ Lett.}\/ {\bf B#1} (19#2) #3}
\def\PRD#1#2#3{{\it Phys.\ Rev.}\/ {\bf D#1} (19#2) #3}
\def\PRL#1#2#3{{\it Phys.\ Rev.\ Lett.}\/ {\bf #1} (19#2) #3}
\def\PRT#1#2#3{{\it Phys.\ Rep.}\/ {\bf#1} (19#2) #3}
\def\MODA#1#2#3{{\it Mod.\ Phys.\ Lett.}\/ {\bf A#1} (19#2) #3}
\def\IJMP#1#2#3{{\it Int.\ J.\ Mod.\ Phys.}\/ {\bf A#1} (19#2) #3}
\def\nuvc#1#2#3{{\it Nuovo Cimento}\/ {\bf #1A} (#2) #3}
\def\RPP#1#2#3{{\it Rept.\ Prog.\ Phys.}\/ {\bf #1} (19#2) #3}
\def\etal{{\it et al\/}}

\bibliographystyle{unsrt}

\end{document}